\documentclass[prd,showpacs,superscriptaddress,twocolumn,
floatfix,preprintnumbers,altaffilletter]{revtex4}
\usepackage{graphicx,url,amssymb,amsmath,longtable,rotating,color,units, 
wasysym,subfigure,epsfig,multirow}
\usepackage{epstopdf}
\usepackage{pdflscape}
\usepackage{makecell}

\newcommand{\drm}{{\rm d}}
\newcommand{\irm}{{\rm i}}
\newcommand{\e}{{\rm e}}
\newcommand{\beq}{\begin{equation}}
\newcommand{\eeq}{\end{equation}}
\newcommand{\bdm}{\begin{displaymath}}
\newcommand{\edm}{\end{displaymath}}

\begin{document}

\title{Towards a first design of a Newtonian-noise cancellation system for Advanced LIGO}
\author{M Coughlin}
\affiliation{Department of Physics, Harvard University, Cambridge, MA 02138, USA}
\author{N Mukund}
\affiliation{Inter-University Centre for Astronomy and Astrophysics (IUCAA), Post Bag 4, Ganeshkhind, Pune 411 007, India}
\author{J Harms}
\affiliation{Universit\`a degli Studi di Urbino ``Carlo Bo'', I-61029 Urbino, Italy}
\affiliation{INFN, Sezione di Firenze, Firenze 50019, Italy}
\author{J Driggers}
\affiliation{LIGO Hanford Observatory, Richland, WA, 99352, USA}
\author{R Adhikari}
\affiliation{Division of Physics, Math, and Astronomy, California Institute of Technology, MS 100-36, Pasadena, CA, 91125, USA}
\author{S Mitra}
\affiliation{Inter-University Centre for Astronomy and Astrophysics (IUCAA), Post Bag 4, Ganeshkhind, Pune 411 007, India}

\begin{abstract}
Newtonian gravitational noise from seismic fields is predicted to be a limiting noise source at low frequency for second generation gravitational-wave detectors. Mitigation of this noise will be achieved by Wiener filtering using arrays of seismometers deployed in the vicinity of all test masses. In this work, we present optimized configurations of seismometer arrays using a variety of simplified models of the seismic field based on seismic observations at LIGO Hanford. The model that best fits the seismic measurements leads to noise reduction limited predominantly by seismometer self-noise. A first simplified design of seismic arrays for Newtonian-noise cancellation at the LIGO sites is presented, which suggests that it will be sufficient to monitor surface displacement inside the buildings. 
\end{abstract}


\maketitle

\section{Introduction}
\label{sec:Intro}

\begin{figure}[t]
 \includegraphics[width=3.3in]{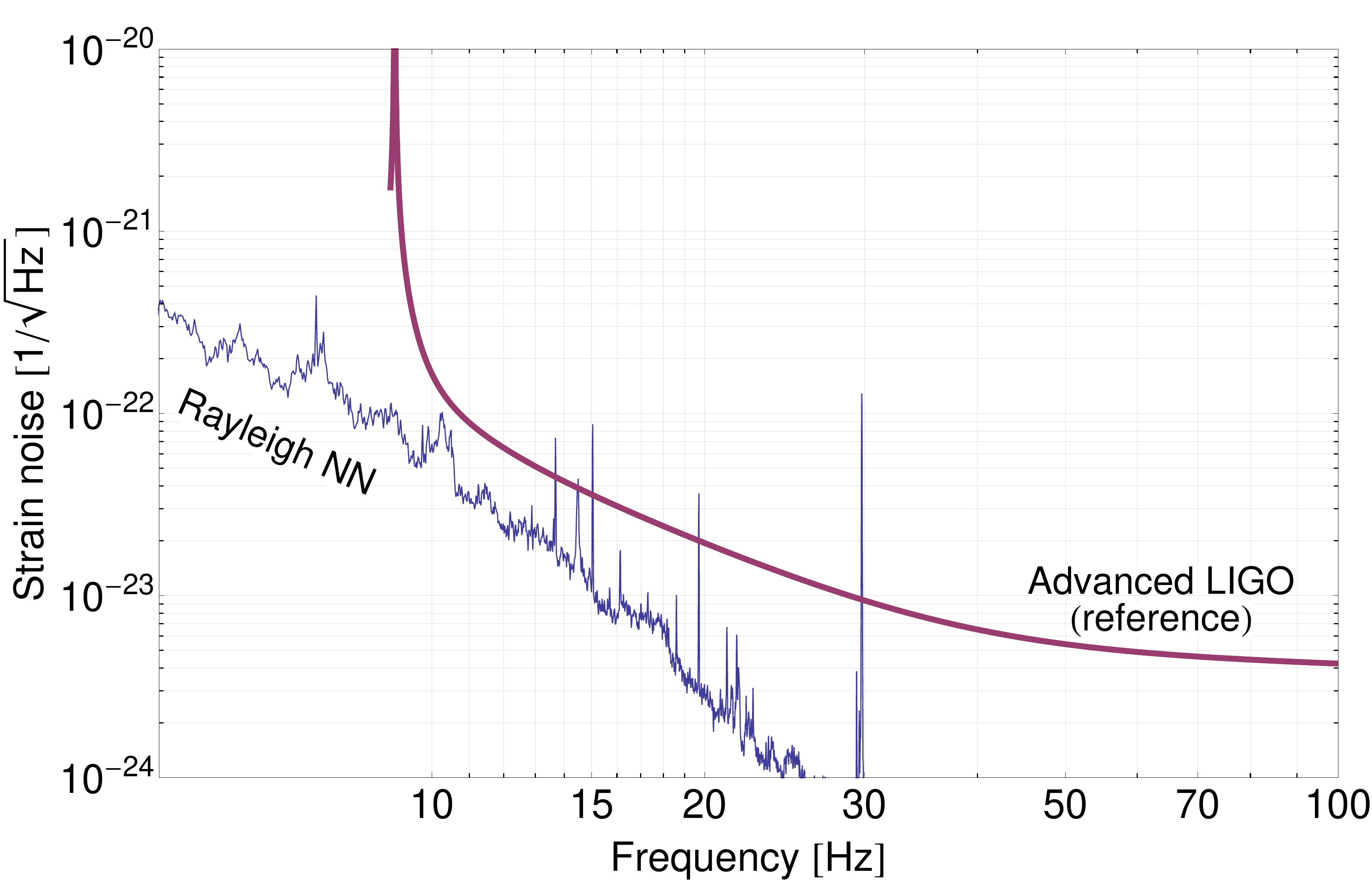}
 \caption{Predicted average Newtonian noise in Advanced LIGO.}
 \label{fig:aLIGONN}
\end{figure}

With the recent detection of gravitational waves produced by a binary black-hole system \cite{AbEA2016a}, efforts to develop novel technology to maximize the sensitivity of the existing gravitational-wave detectors such as Advanced LIGO~\cite{aligo} and Advanced Virgo~\cite{avirgo} are being reinforced. A possible upgrade of the advanced detectors with minimal impact on the detector infrastructures is the cancellation of so-called Newtonian noise (NN) produced by terrestrial gravity fluctuations \cite{Cel2000,BeEA2010,Har2015}. The predicted average Newtonian noise is shown in figure \ref{fig:aLIGONN} together with a reference sensitivity of the Advanced LIGO detectors.
While the average NN spectrum lies below other instrumental noise, it was argued previously that the 90th percentiles will be at the same level of other low-frequency noise, and due to occasional, stronger transients in the seismic field, NN can exceed other noise by a larger factor \cite{DHA2012,Har2015}.

The idea of NN cancellation is to monitor the sources of gravity perturbations, which are generally associated with fluctuating mass density in the vicinity of the test masses. For example, microphones can be used to retrieve information about certain density perturbations in the atmosphere, and seismometers provide information about density perturbations of the ground. Predictions based on a detailed characterization of the LIGO sites show that seismic surface fields give the dominant contribution to NN \cite{DHA2012}. In this case, a NN cancellation scheme can be realized using an array of seismometers deployed at the surface near the test masses \cite{Har2015}. 

In the absence of NN observations, studies of NN cancellation schemes rely on precise modelling. Models of seismic NN have been gradually refined over the past decades \cite{Sau1984,HuTh1998,BeEA1998,Har2015}. Atmospheric NN was identified as a limiting noise source in superconducting gravimeters at mHz frequencies, and noise cancellation successfully implemented using pressure sensors \cite{Neu2010}. However, it is to be expected that atmospheric NN cancellation in future large-scale GW detectors will be significantly more complicated due to a greater variety of atmospheric phenomena that can cause NN \cite{Cre2008}.

The design of a seismic NN cancellation scheme for the advanced detectors needs to address the optimization of the filter used to calculate a NN estimate from seismic data, and also the optimal placement of seismometers around test masses \cite{DHA2012}. The traditional cancellation scheme is based on Wiener filters as was already implemented in feed-forward configuration with respect to other GW detector noise \cite{GiEA2003,DrEA2012,DeEA2012}. Wiener filters are typically calculated from observed correlations between sensors and target channels, but it is also possible to employ models of correlations informed by observations. The latter is advantageous if prior knowledge of the seismic field can be used to constrain the correlation models (such as seismic speeds, or the location of known seismic sources).  

Optimizing the shape of seismic arrays is important for maximizing the efficacy of the noise cancellation. Driggers et al. \cite{DHA2012} first explored schemes to mitigate NN by estimating and subtracting it from the interferometer data stream. They simulated seismic fields with local and non-stationary components, achieving subtraction levels of about a factor of 10 with a finite-impulse response implementation of the Wiener filters. In this paper, we explore using measured seismic fields from an array previously stationed at LIGO Hanford to inform the models for the seismic fields we use to perform array optimization. We include four correlation models fit to the observed data. For all of the models, we use the LHO array data to inform parameters in these models, including seismic speeds. At that point, we use high-dimensional samplers computing optimal arrays for these models by minimizing the expected noise residuals.

The structure of the paper is as follows. We review the formalism of NN and the models used in this analysis in section~\ref{sec:formalism}. We discuss how previous measurements from an array of accelerometers at the LHO site inform our seismic field models in section \ref{sec:lhoarray}. In section \ref{sec:optimization}, we describe how we perform the array optimization. In section \ref{sec:results}, we provide the results of analytic and sampler optimizations for reference seismic fields. Our conclusions are summarized in section \ref{sec:Conclusion}.

\section{Newtonian Noise Cancellation}
\label{sec:formalism}

\subsection{Plane-wave Rayleigh Newtonian noise}
We briefly review Rayleigh-wave NN (see section 3.4.2 in \cite{Har2015} for details). Rayleigh waves are characterized by an evanescent displacement field, which means that displacement amplitude decreases exponentially with distance to the surface. Density perturbations caused by Rayleigh waves can be produced in two ways. First, one needs to consider density changes due to normal surface displacement $\xi_z(\vec r,\omega)$ at frequency $\omega$, where $\vec r=(x,y)$ is the horizontal coordinate vector. Second, Rayleigh waves produce density changes inside the medium. We start with gravity perturbations from pure surface displacement, and focus on homogeneous media in a half-space for simplicity.

If surface displacement is associated with a single Rayleigh wave, then the corresponding gravity acceleration along the horizontal direction $x$ of a test mass at location $\vec r_0$ can be written
\beq
\delta a_x(\vec r_0,\omega)=-2\pi\irm G\rho_0\xi_z(\vec r_0,\omega)\exp(-k h)k_x/k,
\label{eq:surfNN}
\eeq
where $G$ is the gravitational constant, $\rho_0$ the density of the homogeneous medium, $k$ the wavenumber, $h$ the height of the test mass above ground, and the wave vector is $\vec k=(k_x,k_y)$ with $|\vec k|=k$.  Displacement has the same phase at points inside the medium sharing the same horizontal coordinates. This means that additional contributions to the gravity perturbation from density changes inside the medium can be accounted for by multiplying the last equation with a number $\gamma(\nu)$. It depends on the Poisson's ratio $\nu$ of the medium. For Rayleigh waves, the value of this factor is about $\gamma\approx 0.8$ \cite{Har2015}. This means that NN from density perturbations of the medium partially cancel NN from surface displacement.

\subsection{Wiener filtering}
Noise cancellation using Wiener filters exploits correlations between reference data streams and a target data stream to provide a coherent estimate of certain noise contributions to the target sensor \cite{BSH2008}. Wiener filters are the optimal, linear noise-cancellation filters provided that data from all data streams are described by stationary, random processes. In the following, we assume this to be the case.

For seismic NN cancellation, the reference sensors are the $N$ seismometers, and their mutual correlations will be denoted $C_{\rm SS}$ in this paper, which is understood to be a $N\times N$ matrix containing the cross spectral densities of normal surface displacement $C(\xi_z;\vec r_1,\vec r_2,\omega)$ between two seismometers located at $\vec r_1,\,\vec r_2$. The matrix contains seismometer instrumental noise on its diagonal. The target data stream is the output data of the GW detector, but for simplicity, we only consider NN from a specific test mass. In this case, the $N$ seismometers are located in the vicinity of this test mass. The NN spectral density of a test-mass located at $\vec r_0$ is denoted by $C_{\rm TT}\equiv C(\delta a_x;\vec r_0,\omega)$. It is straight-forward to extend the analysis to a cancellation of the total NN from all four test masses.

The correlation between NN and seismometers is denoted by $\vec C_{\rm ST}$, which is a $N$-component vector containing the cross spectral densities $C(\xi_z,\delta a_x;\vec r,\vec r_0,\omega)$ between normal surface displacement observed at $\vec r$ and NN of a test mass at $\vec r_0$. The Wiener filter in frequency space can be written $\vec w=\vec C_{\rm ST}^\top\cdot C_{\rm SS}^{-1}$, where $C_{\rm SS}^{-1}$ is the inverse matrix of $C_{\rm SS}$. The coherent NN estimate is now obtained by multiplying the Wiener filter with the amplitudes of seismic displacement observed by the $N$ seismometers. The matrix $C_{\rm SS}$ and the vector $\vec C_{\rm ST}$ depend on seismometer locations.

Subtracting the coherent NN estimate from the actual test-mass NN, an average relative noise residual $R$ is achieved, which is determined by
\begin{equation}
R = 1 - \frac{\vec{C}_{\rm ST}^\top \cdot C_{\rm SS}^{-1} \cdot \vec{C}_{\rm ST}}{C_{\rm TT}}.
\label{eq:R}
\end{equation}
The cancellation performance can be limited by seismometer instrumental noise, or by the amount of information extracted from the seismic field. We call the first limitation \emph{instrumental}, and the second \emph{geometrical}. For example, a seismometer array with seismometers having distances of order kilometer to each other cannot be used to cancel NN if the seismic wavelength is of order 100\,m independent of the seismometer instrumental noise (unless the seismic field is extremely simple, e.~g.~associated with a single plane Rayleigh wave).

\subsection{Array optimization}
Given a fixed number of seismometers, the optimal array is found by changing seismometer locations and minimizing the noise residual $R$. From eq.~(\ref{eq:R}), one finds that the lowest possible residual with $N$ seismometers is $R_{\min}(\omega)=1/(N\sigma(\omega)^2)$ with $\sigma(\omega)$ being the signal-to-noise ratio of the seismometers assumed to be equal for all sensors. This residual can only be achieved when geometrical limitations are overcome, which is not always possible even for an infinite number of seismometers (see Section~\ref{subsec:arrays}). 

We could now construct specific plane-wave compositions of the seismic field and use eq.~(\ref{eq:surfNN}) to evaluate and minimize the noise residual in eq.~(\ref{eq:R}). While this has some merit for analytical understanding of the optimization, the goal of this paper is to present a method to use observable quantities of the seismic field to model NN and its cancellation. Seismic fields can show great complexity for example due to seismic scattering, or presence of local sources, in which case the plane-wave model is impractical. The key here is to realize that all correlation functions required for calculating the noise residual can be expressed in terms of the two-point spatial correlations $C(\xi_z;\vec r_1,\vec r_2,\omega)$ of the Rayleigh field:
\begin{widetext}
\begin{equation}
\begin{split}
C(\delta a_x;\vec\rho_0,\omega) &= (2 \pi G \rho_0 \gamma(\nu))^2 \int \int \frac{\drm^2 k}{(2 \pi)^2} \frac{\drm^2 k'}{(2 \pi)^2} S(\xi_z;\vec{k},\vec{k}',\omega)
\frac{k_x}{k} \frac{k'_x}{k'} \e^{-h k} \e^{-h k'} \e^{\irm \vec r_0 \cdot (\vec{k} - \vec{k}')}, \\
C(\delta a_x;\vec r_0,\omega) &= (G \rho_0 \gamma(\nu))^2 \int \int \drm^2r \,\drm^2r' C(\xi_z;\vec r,\vec r\,',\omega)
\mathcal K(\vec r,\vec r_0)\mathcal K(\vec r\,',\vec r_0),\\[0.3cm]
C(\xi_z,\delta a_x;\vec r_0,\vec r,\omega) &= - 2 \pi \irm G \rho_0 \gamma(\nu) \int \int \frac{\drm^2 k}{(2 \pi)^2} \frac{\drm^2 k'}{(2 \pi)^2} S(\xi_z;\vec{k},\vec{k}',\omega) \frac{k_x}{k} \e^{-h k} \e^{\irm (\vec r_0 \cdot \vec{k} - \vec r \cdot \vec{k}')},\\
C(\xi_z,\delta a_x; \vec r_0,\vec r,\omega) &= G \rho_0 \gamma(\nu) \int \drm^2r' C(\xi_z;\vec r,\vec r\,',\omega) \mathcal K(\vec r\,',\vec r_0).
\end{split}
\label{eq:correlations}
\end{equation}
\end{widetext}
with
\beq
\begin{split}
S(\xi_z;\vec k,\vec k',\omega)&\equiv \int \int \drm^2r \,\drm^2r' C(\xi_z;\vec r,\vec r\,',\omega)\e^{-\irm(\vec r \cdot \vec k - \vec r\,' \cdot\vec k')}\\
\mathcal K(\vec r_1,\vec r_2)&\equiv \frac{x_1 - x_2}{(h^2 + |\vec r_1-\vec r_2|^2)^{3/2}}.
\end{split}
\eeq
Simplified versions of these equations for a homogeneous field, $C(\xi_z;\vec r,\vec r\,',\omega)=C(\xi_z;\vec r-\vec r\,',\omega)$, or a homogeneous and isotropic field, $C(\xi_z;\vec r,\vec r\,',\omega)=C(\xi_z;|\vec r-\vec r\,'|,\omega)$, are presented in \cite{Har2015}. 

The first and third line in eq.~(\ref{eq:correlations}) follow from eq.~(\ref{eq:surfNN}) by calculating expectation values of products of displacement amplitudes at different wavenumbers (note that we do not impose the condition $k=\omega/c$, which holds for waves propagating at speed $c$). The second and fourth line provide the desired link between seismic observation and NN modelling. It is important that these equations only require seismic correlations and no other prior knowledge of the seismic field such as seismic speeds. Of course, the simplest analytical models of $C(\xi_z;\vec r,\vec r\,',\omega)$ that we can construct will have additional parameters, and seismic wavelength or equivalently seismic speed may be one of them. We are now prepared to either directly use measured correlations between seismometers, or models fit to observed correlations, to optimize seismometer arrays for NN cancellation. Consequently, the path towards a NN cancellation scheme includes a detailed site characterization to measure two-point spatial correlations of the seismic field near all test masses.

\section{LIGO Hanford Array}
\label{sec:lhoarray}

\begin{figure}[ht!]
\includegraphics[width=3.3in]{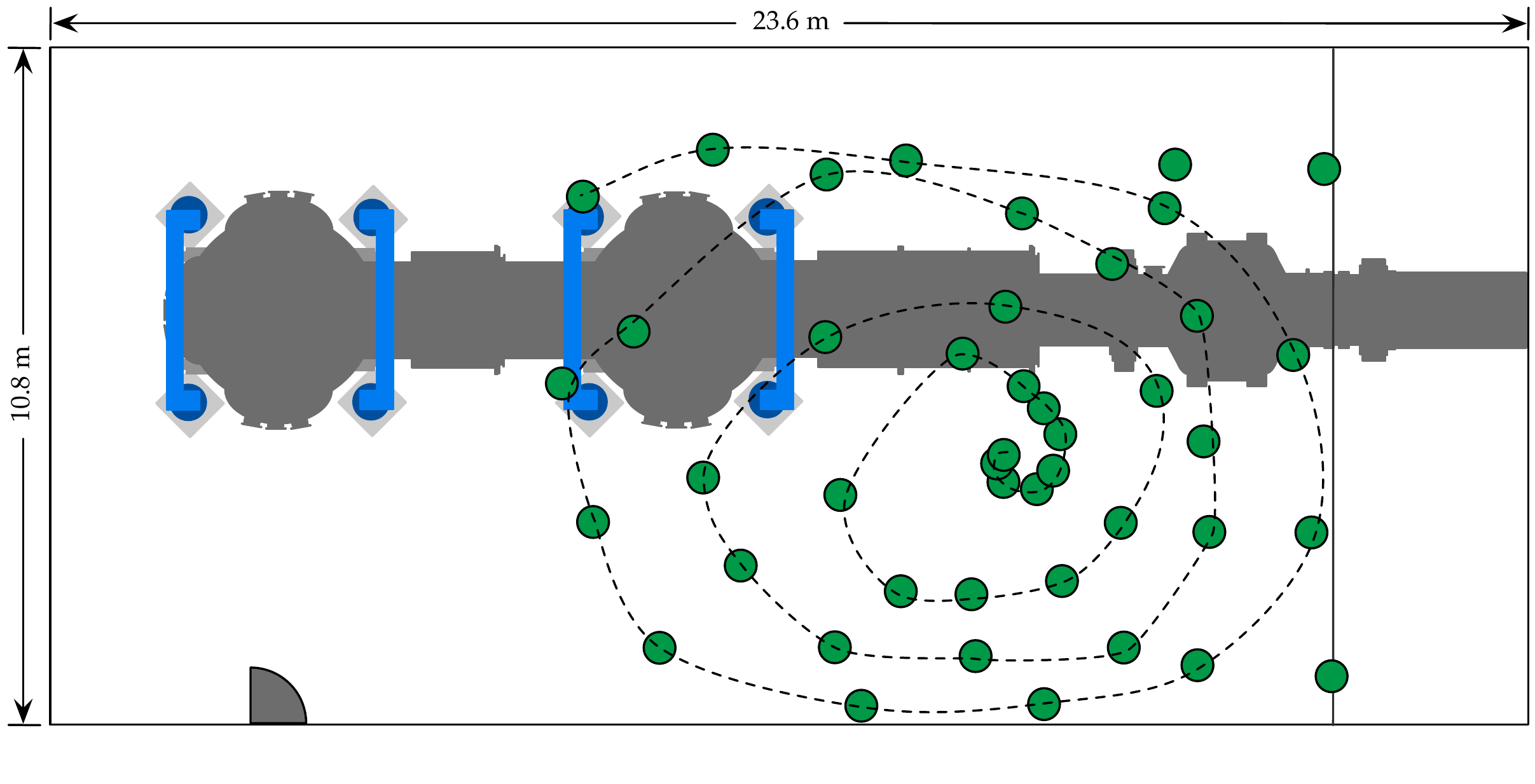}
\caption{Layout of the instrument floor at the Y-end station of the LIGO Hanford Observatory. Green circles indicate placement of accelerometers in the vicinity of the vacuum enclosure. Dotted line indicates the spiral shape of the array. Due to ongoing work at the time, the array is not centered around the vacuum chamber that holds the test mass mirror.}
\label{fig:array2012}
\end{figure}

In 2012, from April through November, an array of 44 Wilcoxon Research 731-207 accelerometers~\cite{Wilcoxon731207} was deployed at an end station building at LIGO Hanford.  Figure~\ref{fig:array2012} shows the locations of accelerometers in the vicinity of the vacuum enclosure.  Each accelerometer's signal was conditioned, including amplification of a factor of 100, before being acquired digitally and saved at 1024\,Hz sampling rate. The array configuration followed as closely as possible the shape of a spiral since it has favorable properties for analyzing seismic fields by naturally suppressing aliasing effects when performing analyses in spatial Fourier domain \cite{Wat2005}. 

\begin{figure}[t]
 \includegraphics[width=3in]{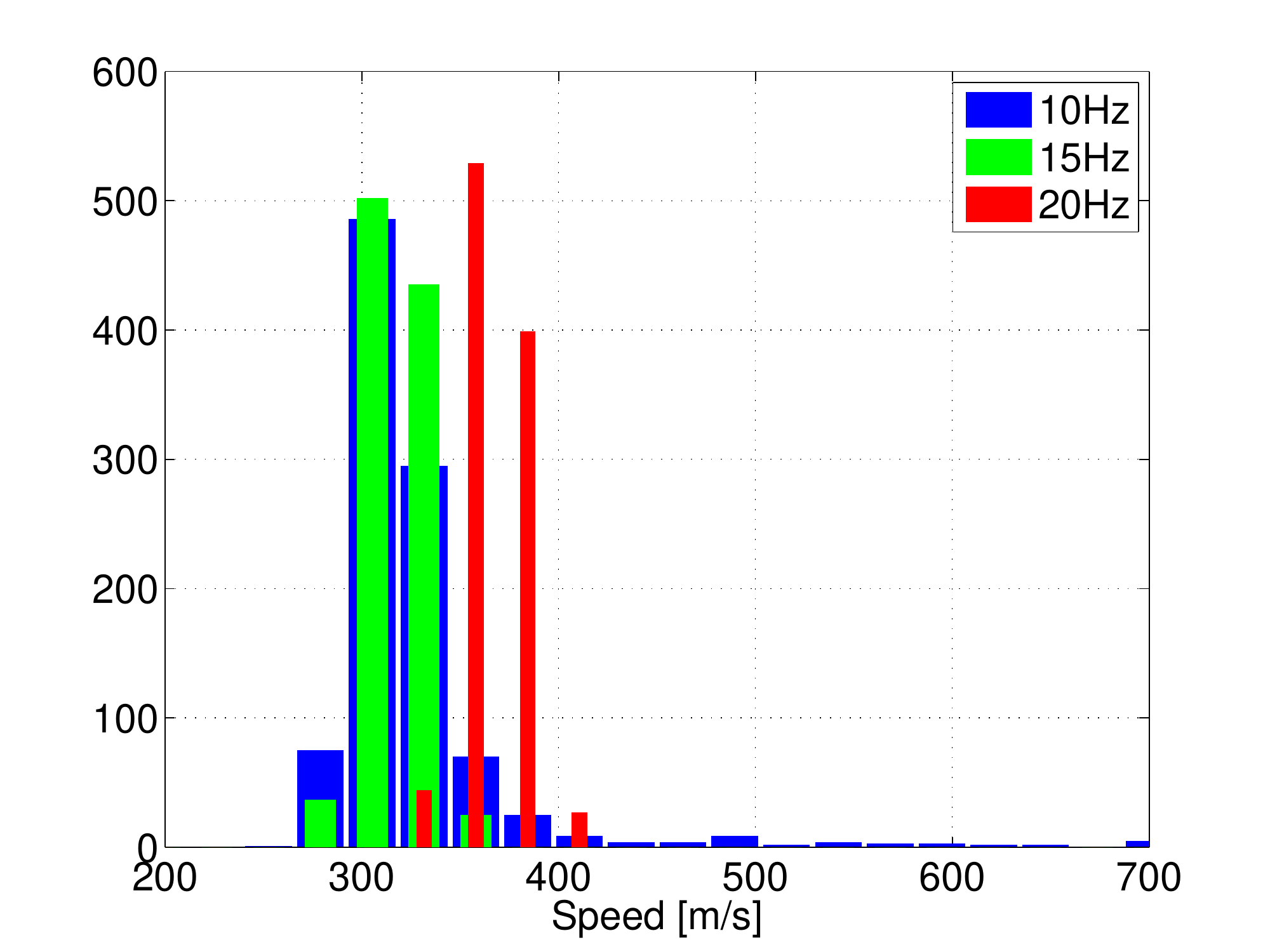}
 \caption{Histograms of seismic speed measurements using LHO array data at 10, 15, and 20\,Hz.}
 \label{fig:speeds}
\end{figure}

Two analyses of the array data are presented in the following, which are directly relevant to NN modelling. The first analysis is a measurement of Rayleigh-wave speed in the frequency range 10 -- 20\,Hz, which is chosen to cover the band likely to be limited by NN. The method used here is to decompose the seismic field into plane harmonics (see section 3.6.3 of \cite{Har2015}), and collect the phase speeds associated with the maximum-amplitude component over a period of a week. Longer stretches of data are not required for the purpose of this paper since seismic spectra between 10 -- 20\,Hz show a high level of stationarity at the LIGO sites. Figure \ref{fig:speeds} shows that mean seismic speed at 10\,Hz and 15\,Hz is around 300\,m/s consistent with previous site-characterization measurements \cite{ScEA2000}, while at 20\,Hz measured speeds around 380\,m/s are higher than observed previously. This anomalous dispersion can be explained by the concrete slab of the laboratory building, which has greater effect at short seismic wavelengths. A similar result can be expected for the LIGO Livingston site, where Rayleigh-wave speeds in this frequency range estimated from field measurements are around 200\,m/s \cite{HaOR2011}. While the Rayleigh-wave speed is not required to evaluate the integrals in eq.~(\ref{eq:correlations}), it can be used to define simple correlation models as shown in section \ref{subsec:models}. 

The speed histograms also indicate that seismic noise is dominated by Rayleigh waves. The high-speed tail of the distribution, which is potentially associated with seismic body waves, has very small values. This is a very important conclusion for NN modelling and greatly impact plans for NN cancellation. It remains to be seen of course, if the situation is similar at Livingston and other stations of the Hanford detector.

\begin{figure}[t]
 \includegraphics[width=3.3in]{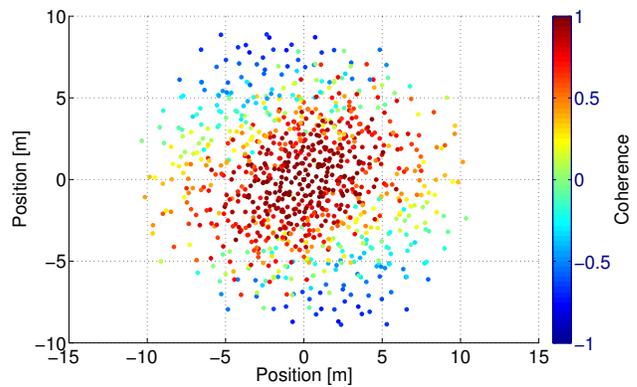}
 \caption{Measured correlation function $C(\xi_z;\vec\varrho,\omega)$ for the LIGO Hanford site at 15\,Hz.}
 \label{fig:corrLHO}
\end{figure}

The second analysis is a measurement of seismic correlations $C(\xi_z;\vec\varrho_1,\vec\varrho_2,\omega)$ between all 29 accelerometers used for this study. Only 29 out of 44 accelerometers were used due to the others being mislabeled in the data acquisition. Figure \ref{fig:corrLHO} shows values of $C(\xi_z;\vec\varrho,\omega)$ at 15\,Hz with $\vec\varrho=\vec\varrho_2-\vec\varrho_1$ and $\vec\varrho=\vec\varrho_1-\vec\varrho_2$. If the seismic field were homogeneous, then $C(\xi_z;\vec\varrho_1,\vec\varrho_2,\omega)=C(\xi_z;\vec\varrho_2-\vec\varrho_1,\omega)$, and the function outlined by the scatter plot would be smooth. However, there are parts in this plot where high correlation values exist very close to low correlation values. So the observed $C(\xi_z;\vec\varrho_2-\vec\varrho_1,\omega)$ is not smooth, and therefore the seismic field is inhomogeneous. The most likely cause of inhomogeneities are local sources, which affect some seismometers of the array more strongly than others altering correlations throughout the array. Nonetheless, in this paper, we will only fit homogeneous models to the scatter plot, assuming that we can understand the result as a smoothly changing correlation function $C(\xi_z;\vec\varrho_2-\vec\varrho_1,\omega)$ perturbed by a sufficiently small number of outliers.

If inhomogeneities were more pronounced, then NN modelling and array optimization must be directly based on eq.~(\ref{eq:correlations}) valid for inhomogeneous fields. Since it is very hard to conceive a model of $C(\xi_z;\vec\varrho_1,\vec\varrho_2,\omega)$ that could be fit to observed inhomogeneous correlations, the better strategy may be to calculate the integrals numerically over observed values, possibly using interpolation techniques. These schemes were tested in preparation of this article, but it was found that array optimization based on numerical integration requires correlation measurements between a larger number of seismometers. The crucial difference to optimization in homogeneous fields is that with $N$ seismometers, for each seismometer, the function $C(\xi_z;\vec\varrho_1,\vec\varrho_k,\omega)$ with $k=1,\ldots, N$ is only sampled $N$ times, while an estimate of the homogeneous correlation $C(\xi_z;\vec\varrho_2-\vec\varrho_1,\omega)$ is calculated from $N\times N$ samples. So given a certain precision goal, which is determined by the targeted noise-suppression factor, a much larger number of seismometers is required in the case of inhomogeneous fields compared to homogeneous fields. If in the case of significant inhomogeneities, which would be directly visible in correlation maps or spatial Fourier spectra, the seismic characterization cannot be done with a sufficient number of seismometers to make possible a calculation of fully optimized arrays, the alternative would be to take the homogeneous optimization as starting point, and to try to identify the causes of the inhomogeneities, and to improve cancellation performance based on additional modelling work. Consequently, the level of homogeneity of the seismic field is one of the most important properties concerning NN cancellation.

\section{Array Optimization}
\label{sec:optimization}

There exist different methods to find the optimal array, all of which are based on minimizing eq.~(\ref{eq:R}). The method used in \cite{Har2015} first applies analytic transformations of the equation to determine the optimal array as zeros of a new set of equations. Alternatively, one can directly find the minimal residual by applying generic high-dimensional sampling algorithms such as nested sampling, Metropolis Hastings MCMC, or particle swarm optimization (the last was used in \cite{DHA2012}). For each instance of array generated by these algorithms, we compute either the analytic or measured seismometer correlations, denoted by $C_{\rm SS}$, thereby $C_{\rm SN}$ using eq.~(\ref{eq:correlations}) and hence estimate the cost function. For the case of $N$ sensors the cost function involves a parameter space with dimensionality $D=2N$ and is characterized by numerous local minima. The ideal algorithm would be the one capable of optimizing sensor locations simultaneously and obtain solutions close to global minima within a reasonable execution time.  \\

The results in this paper are calculated with a Metropolis Hastings MCMC algorithm implementing adaptive simulated annealing (SA), which statistically guarantees solutions close to global minima \cite{Kirkpatrick671,Ingber_ASA}. SA is a probabilistic scheme constructed to begin as a random exploration of the parameter space and later progress with time towards a greedy kind of search. The search begins with non negligible  chance of accepting bad solutions whose probability slowly decreases as time progresses. This feature helps evade getting stuck in local minima leading to a final near optimal sensor array configuration. \\

 In our implementation, we perform parallel analysis at each iteration by executing a MATLAB implementation of adaptive SA \cite{MATLAB_2013} on different computing cores and selecting the best configuration for  use in further iterations.  Here the annealing schedule and the process of generating newer samples  get adaptively modified so as  to achieve accelerated convergence. The iterative process is repeated until the algorithm's stopping criterion is reached, and thereafter the array with the smallest residual is taken to be the optimal array. The success of the optimization depends on careful selection of the probability distribution $g_{T}(x)$ used to span the multi-dimensional parameter space, acceptance criteria $h(\Delta{E})$  used to select the newer points and the temperature $(T)$ or the annealing schedule, which determines the level of vigorousness associated with the search.  \\

Acceptance function is given by $h(\Delta{E}) = 1/(1+\Delta{E}/T)$ where $\Delta{E}$ is the difference in cost function between the new and the previous state. We represent the state space probability distribution using a Cauchy distribution $g(\Delta{x}) = T/(\Delta{x}^{2} + T^{2})^{(D+1)/2}$. Temperature at future time steps is calculated using an exponential annealing schedule  $T_{i}(k) = T_{0i}\; \exp (c_{i} k^{1/D} ) $ with $c_{i} = m_i \; \exp(-n_i/D)$. Here $m_{i}$ and $n_{i}$  act as free parameters that are used to fine tune the optimization process. The variation in sensitivities of different parameters during the course of the search is addressed by reannealing or rescaling the time-steps after every user-defined number of accepted states. This essentially  decreases or increases the range for those parameters, which show more or less sensitivity at the current annealing schedule. Accuracy of the search after each SA iteration is further enhanced by performing a polling based pattern search \cite{MATLAB_2013}, which is well suited for carrying out local minimization when the problem at hand is ill-defined and non-differentiable.  \\

We make a number of assumptions in the following analysis. First, since seismic fields are highly stationary in the 10 -- 20\,Hz range at the LIGO sites \cite{Har2015}, arrays are optimized towards maximal reduction of a stationary NN background. It is possible that rare strong transients in the 10 -- 20\,Hz range are better suppressed by arrays with different configurations. Second, we assume a signal-to-noise ratio (SNR) of 100 for the seismic sensors. In low-noise environments such as underground sites even the most sensitive seismometers do not necessarily achieve SNRs of 100 in the NN band, but at the existing sites of the LIGO and Virgo detectors with seismic spectra being a few orders of magnitude above the global low-noise model \cite{Har2015}, the most sensitive instruments in the NN band such as GS-13, STS-2, or T240 have SNRs of a few 1000, and relatively inexpensive sensors already have SNRs of a few 100. Third, we choose for the test mass to be suspended 1.5\,m above ground, which is approximately the height of the LIGO test masses. It should be noted though that the test-mass height has no effect on optimal arrays for cancellation of NN from plane-wave Rayleigh fields. In our simulation, we search for optimal sensor locations within a 100\,m $\times$ 100\,m surface area with a test mass at its center. This is conservative, as this is larger than the area from which interesting NN contributions are expected \cite{HaEA2009a}. 

\section{Results}
\label{sec:results}

The structure of this section is as follows. We present the models we use to fit to the measured LHO array seismic field from section~\ref{sec:lhoarray} in subsection~\ref{subsec:models}. We then use those models in subsection~\ref{subsec:arrays} to find optimal seismic arrays for each of the models presented. 

\subsection{Models}
\label{subsec:models}
\begin{figure*}[t]
 \includegraphics[width=2.2in]{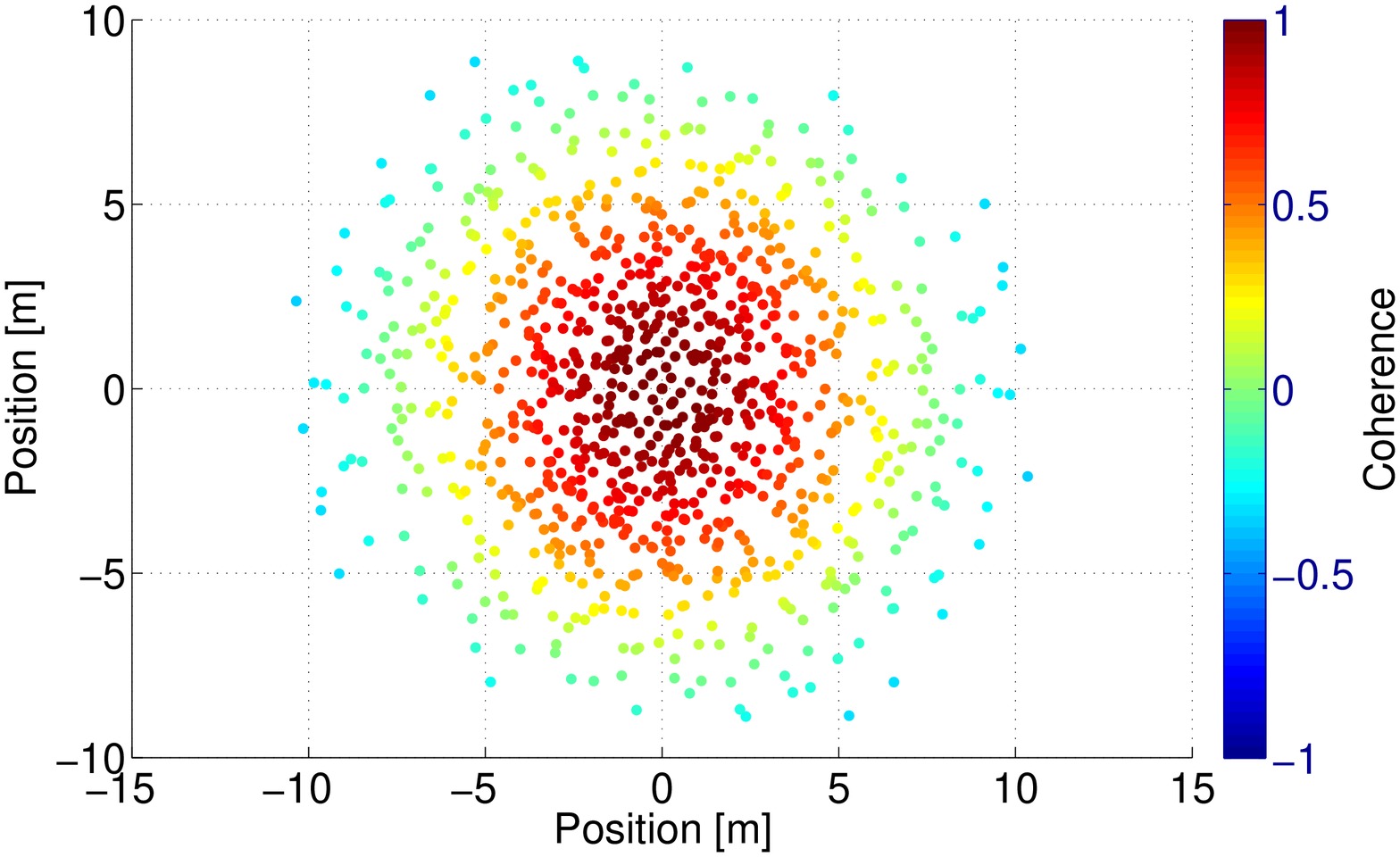}
 \includegraphics[width=2.2in]{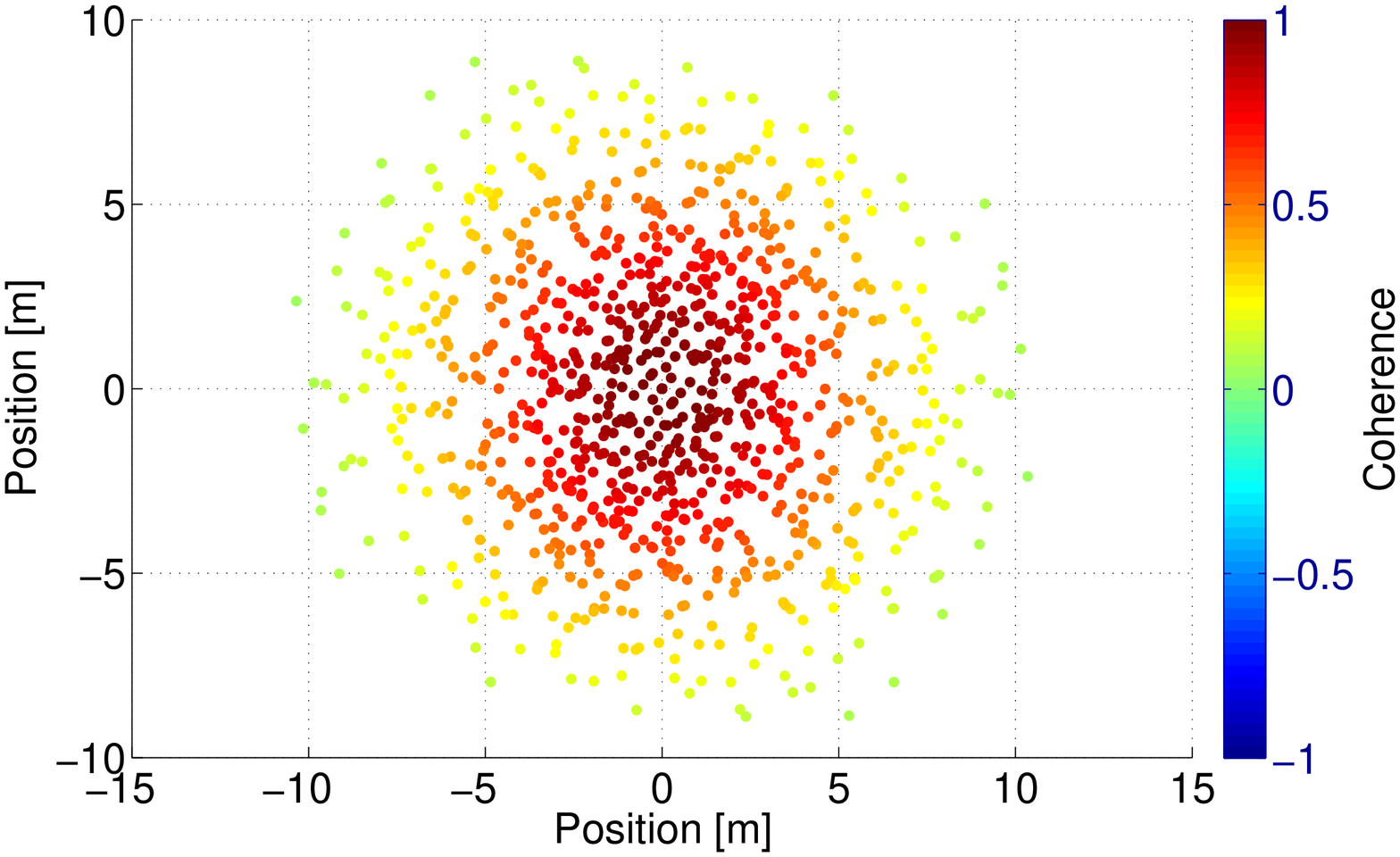}
 \includegraphics[width=2.2in]{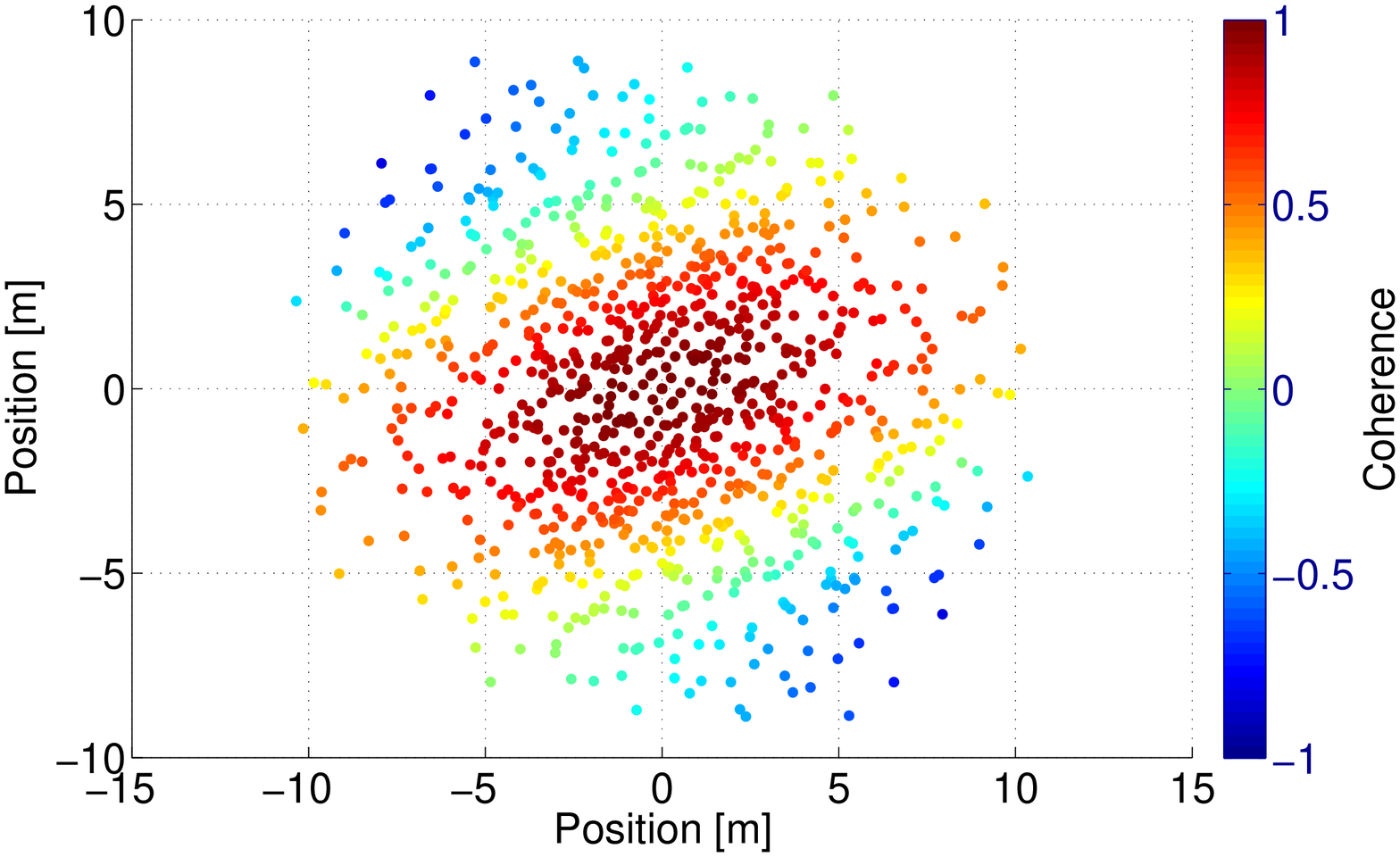}
 \caption{From the left to right are the IPW, Gaussian, and APW models for the seismic fields.}
 \label{fig:maps}
\end{figure*}

We solve the integrals of eq.~(\ref{eq:correlations}) for various models of the seismic correlation $C_{\rm SS}$, including an isotropic plane-wave (IPW) model, which takes the form of a Bessel function \cite{DHA2012}, an isotropic Gaussian model, an anisotropic plane-wave (APW) model, and the case of a single plane wave (SPW).

Explicit expressions for $C_{\rm SS}$ exist for the IPW model, $C_{\rm SS} = J_0 (k |\vec{r}_i - \vec{r}_j|)$, the Gaussian model, $C_{\rm SS} = \exp(-|\vec{r}_i - \vec{r}_j|^2/\sigma^2)$, and the SPW model, $C_{\rm SS} = \cos(\vec k\cdot (\vec r_i - \vec r_j))$. In the APW case, the seismic correlations are calculated numerically by averaging plane-wave contributions over propagation directions with a direction dependent Gaussian weight $A(\phi) = \exp(-(\phi-\phi_0)^2/\delta^2)$, where $\phi$ is the polar angle quantifying the direction of a Rayleigh wave. 

Using the LHO array data presented in section~\ref{sec:lhoarray}, we perform fits of the above models to the measured $C_{\rm SS}$. Taking 300\,m/s as the seismic speed, we find that at 15\,Hz, the length of seismic waves is $\lambda=20$\,m, which directly determines the wavenumbers of the IPW, SPW, and APW models. For the anisotropic model, we find best fit parameters of $\phi_0 = 110^\circ$ and $\delta = 52^\circ$. While these values do not have a direct impact on NN cancellation (cancellation is weakly affected by anisotropy), it means that the absolute NN level is lower compared to the isotropic case. This is because seismic waves are propagating almost perpendicularly to the interferometer arm; see eq.~(\ref{eq:surfNN}). In practice, there is an indirect effect on NN cancellation, since correlation between seismometers and GW channel is reduced when NN is weaker, which could mean that more data are required to calculate the Wiener filter.

The SPW and Gaussian models are not based on independent fits to observed correlations, but are based on the above best-fit parameter values. The reason for this is that both models are poor representations of observed correlations, but serve as lower (SPW) and upper (Gaussian) bound of achievable noise residuals. 

The APW value of $\phi_0$ is also used for the propagation direction of the SPW model, which, together with the wavelength parameter from above, determines the wave vector $\vec k$. The Gaussian model parameter is defined as $\sigma = \lambda / \pi$, which yields $\sigma = 6.5$\,m. In this way, the Gaussian model is at least a good fit to observed short-range correlations, while neglecting deviations at larger sensor separations. The Gaussian model is an isotropic, homogeneous correlation model, but it implies a significant impact from local seismic sources, which can also include scattering. There is no underlying plane-wave representation of the seismic field since this would necessarily give rise to a IPW or APW type model. The IPW, APW, and Gaussian models evaluated for the sensor pairs of the LHO array are shown in figure \ref{fig:maps}.

\subsection{Optimal Arrays}
\label{subsec:arrays}

\begin{figure*}[ht!]
\includegraphics[width=3.5in]{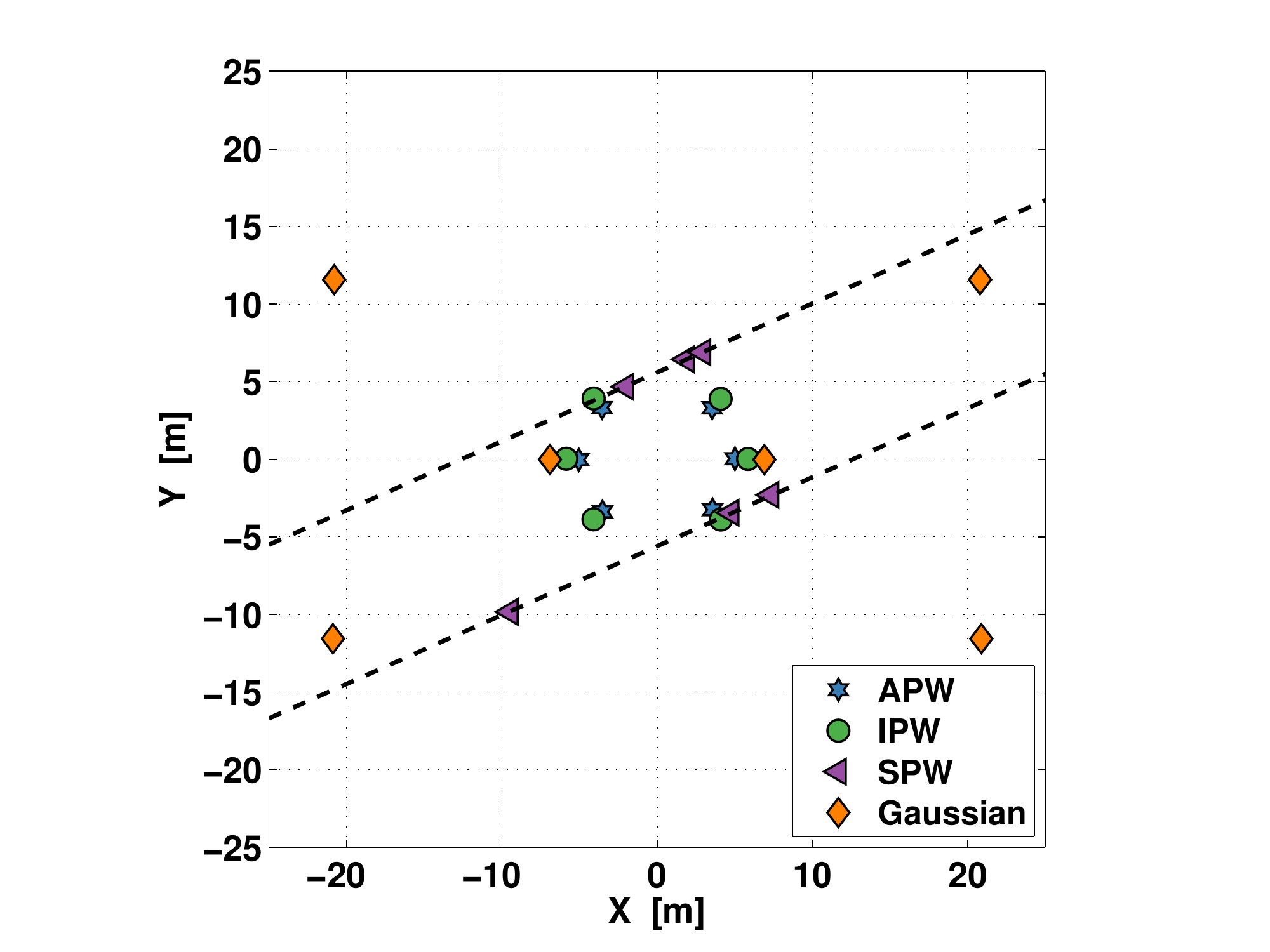}
\includegraphics[width=3.5in]{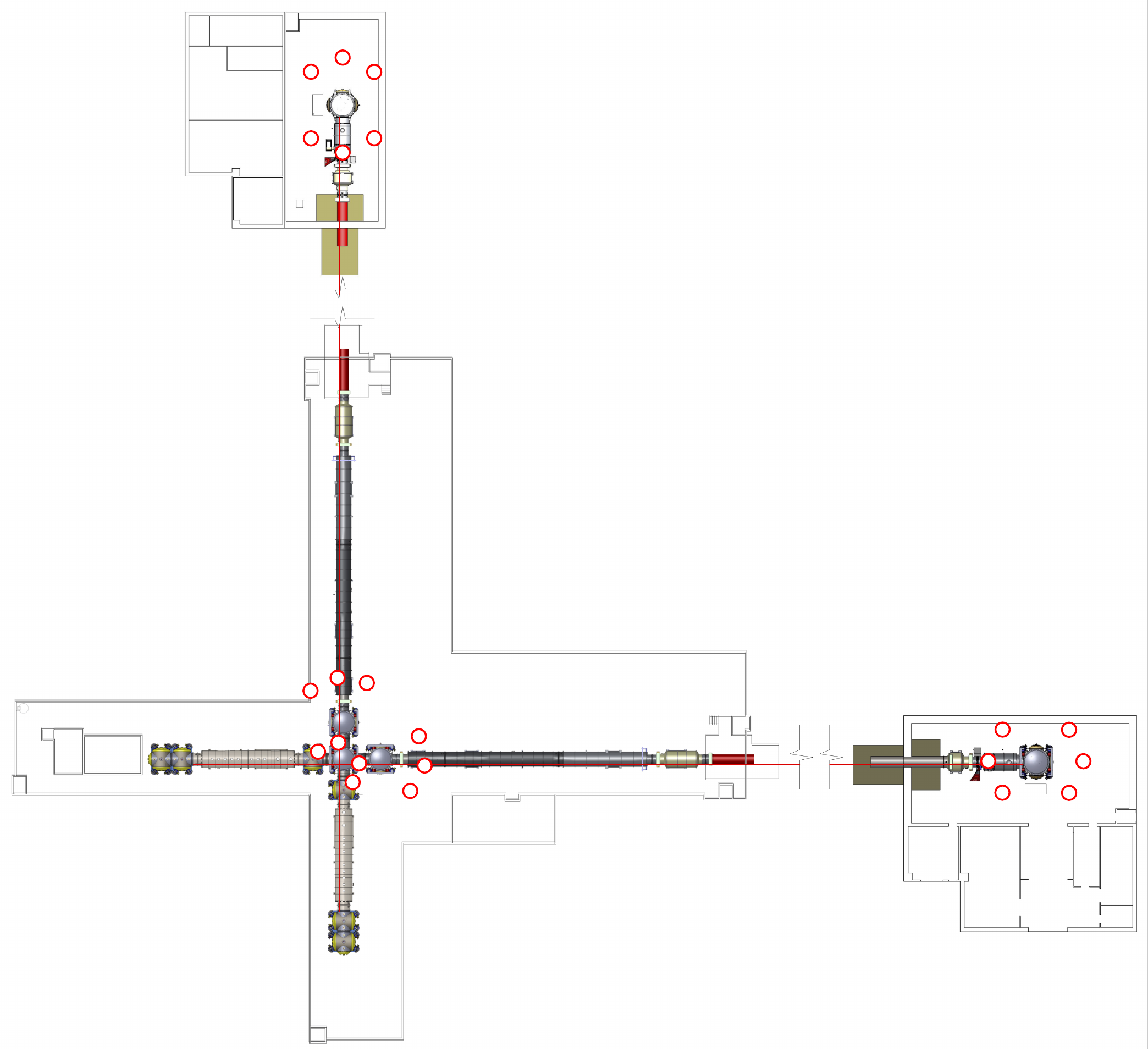}
\caption{On the left are the locations of sensors resulting from numerically minimizing the subtraction residual for the IPW, APW, SPW, and Gaussian correlation functions. The dashed lines indicate that optimal locations of seismometers for the SPW can lie anywhere on these lines. On the right is an example arrangement of seismometers at both end stations and the corner stations, optimized for cancellation at 15\,Hz. The end station placement is based on the IPW model discussed in the text. The corner station arrangement is also based on the IPW model, but this time simultaneously minimizing noise residuals from both test masses.}
\label{fig:lhoarray}
\end{figure*}

The structure of the optimal array analysis is as follows. For each of the models presented in section \ref{subsec:models}, we use the optimization scheme presented in section \ref{sec:optimization} to minimize the NN residuals, given by eq.~(\ref{eq:R}), at 15\,Hz. We perform the optimization for each of these cases for arrays containing from 1 to 20 sensors, which allows for the exploration of how many sensors are necessary to achieve a required subtraction.

As discussed above, the optimization analysis is performed minimizing residuals at 15\,Hz, corresponding to a wavelength of $\lambda = 20$\,m assuming a seismic speed of 300\,m/s, and a seismometer SNR of 100. On the left of figure \ref{fig:lhoarray}, we show optimized seismometer arrays with six sensors for all correlation models. The key features of all arrays optimized for a single test mass are that they are symmetric with respect to the $x$-axis pointing in the direction of the interferometer arm, with their distance to the test mass located at the center of the plot proportional to the wavelength of interest. This optimal distance depends also on the sensor SNR. 

The IPW and APW models yield similar optimal arrays with distances between seismometers and the test mass being slightly shorter for the APW model. This greatly facilitates the design of NN cancellation systems, since anisotropies of the field play a minor role. For the SPW model, there are no distinct solutions for optimal seismometer locations. Instead, seismometers can be deployed anywhere on the two dashed lines and lines parallel to this at larger distances separated by half a wavelength. Also the Gaussian model produces a qualitatively different optimal array, which is further evidence of the fact that there is no underlying plane-wave representation of the seismic field. 

For all correlation models, seismometer are placed away from the test mass by a significant fraction of the seismic wavelength. This is because a sensor located directly under the test mass has vanishing correlation with NN, which means that a sensor not directly under but very close to the test mass must have extremely high SNR to show significant correlation with NN. Therefore, given a finite SNR, optimal array configurations with any number of sensors place seismometers sufficiently far away from the test mass according to their SNR, but not so far that correlation with test-mass NN becomes too small for effective cancellation. This means that the sensors can be placed inside of the building, which is beneficial for practical reasons since measurements with outdoor sensors can be affected by weather even if well protected inside vaults. 

Interestingly, in the case of the IPW and APW models, the 6-sensor optimal arrays also represent the optimal locations of arrays with a higher number of sensors. In other words, the optimization algorithm yields solutions where seismometers are placed on top of each other. In practice, this can be realized by placing seismometers as close to each other as possible, or to conclude that instead of adding new seismometers to a 6-sensor array, seismometers with higher SNR should be used. For the same reason, the 6-sensor configuration is independent of the seismometer SNR, which is not true for optimal arrays with less than 6 sensors where seismometers with higher SNR are positioned closer to the test mass.

In the right plot of figure \ref{fig:lhoarray}, optimized arrays are again shown for the IPW model. The optimal array at the corner station has a different configuration since seismometers are used to cancel NN from both test masses simultaneously. The effect is greatest on placement of seismometers that are close to both test masses. According to these results, seismic arrays for NN cancellation are located completely inside the buildings. 

\begin{figure*}[ht!]
\includegraphics[width=3.25in]{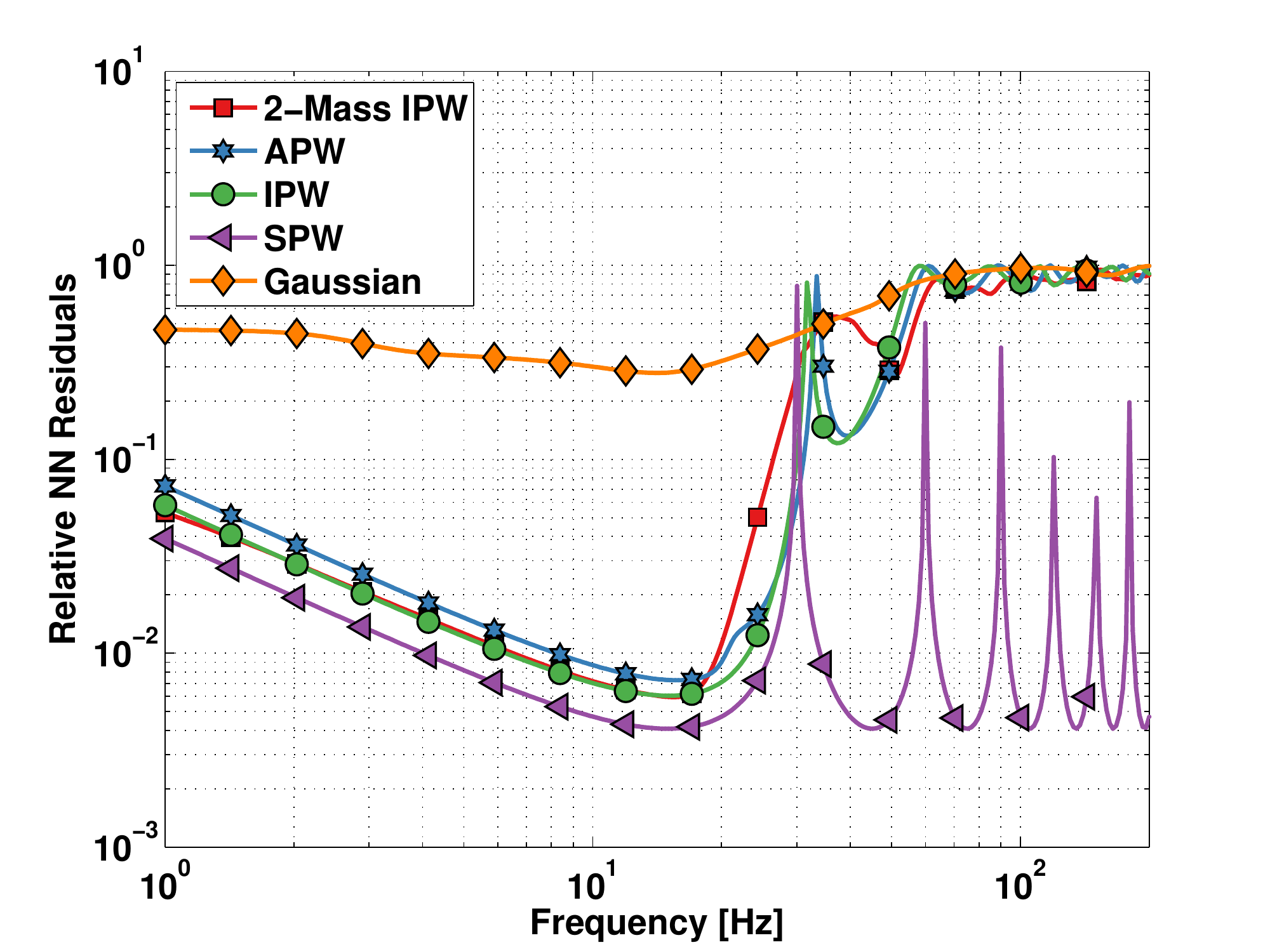}
\includegraphics[width=3.3in]{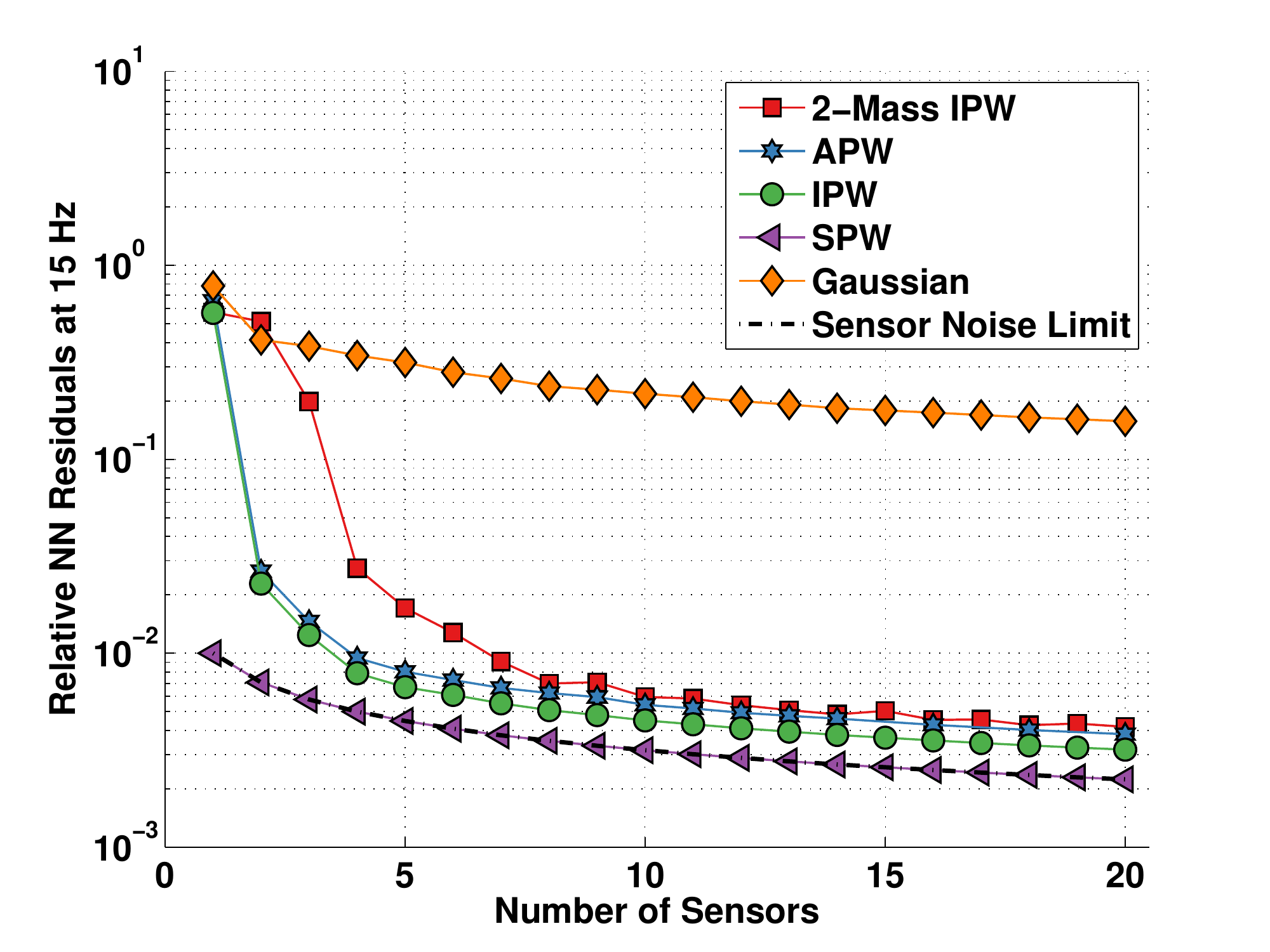}
\caption{The left plot shows the subtraction residual vs. frequency for the optimal arrays based on the various correlation models. The ``2-mass'' IPW model is for simultaneous minimization of residuals from both test masses at the corner station with a 10 seismometer array. All other curves are for the 6 seismometer arrays shown in Figure \ref{fig:lhoarray}. On the right is the noise reduction at 15\,Hz as a function of the number of sensors. For the 6 seismometer arrays, a reduction by approximately a factor of 3.5 is achieved for Gaussian, 150 for IPW, 140 for APW, and 250 for SPW. The last is equal to the theoretical maximum of noise suppression given by SNR $\times \sqrt{N}$.}
\label{fig:subtraction}
\end{figure*}

Figure \ref{fig:subtraction} shows subtraction residuals. It is assumed that seismometers have a frequency-independent SNR of 100. In the left plot, residual noise spectra relative to the unsuppressed NN spectra are shown using 6-sensor arrays in the case of the single test-mass cancellation, and 10 sensors for the 2-mass cancellation. The Gaussian residuals are consistently high over the entire frequency range supporting its role as worst-case scenario. In all spectra, residuals increase towards lower frequencies. It is a consequence of the fact that the ability of an array to extract information from the seismic field as relevant to NN cancellation decreases if the separation between sensors becomes significantly smaller than the seismic wavelength. At high frequencies, wavelengths become smaller than distances between sensors decreasing correlations between seismometers and NN, again degrading the cancellation performance. 

For the APW and IPW models with one or two test masses, increasing sensor SNR and using the same seismometer locations, residuals at frequencies below the minimum at 15\,Hz decrease accordingly, while the residuals above 15\,Hz do not change significantly. Adding sensors at larger distances to the test mass would not significantly change residuals above 15\,Hz, but lead to further modest decrease of residuals below 15\,Hz. Adding sensors at smaller distances to the test mass would not have a great impact on residuals below 15\,Hz, but would extend the frequency band with good cancellation performance to higher frequencies. In summary, if it is necessary to improve residuals at frequencies below the minimal residual, then the best option is to use seismometers with higher SNR. If it is necessary to lower residuals above the minimum, then the best option is to add further seismometers at shorter distance to the test mass. With respect to the SPW model, adding sensors at smaller and larger distances to the test mass or increasing SNR improve residuals at all frequencies, and potentially eliminate the oscillatory behavior of SPW residuals at higher frequencies. For the Gaussian model, neither adding sensors nor increasing the SNR has a significant effect on noise residuals, unless the extended array is reoptimized for the new SNR or number of seismometers.

Scaling the seismometer locations relative to the test mass by a constant factor, the residual noise spectra of the plane-wave models with a single test mass are simply shifted towards higher or lower frequencies otherwise maintaining absolute values and shapes. This is not the case for the Gaussian model where NN residuals depend on the test-mass height (this dependence cancels out in all plane-wave models), and for the 2-mass IPW model where the distance between test masses introduces another reference length.

The right plot in figure \ref{fig:subtraction} shows noise residuals at 15\,Hz as a function of number of sensors. The residuals can be compared with the absolute sensor-noise limit $1/({\rm SNR}\sqrt{N})$ shown as dashed line, which marks the minimal residuals that can possibly be achieved with $N$ sensors independent of the seismic correlation model. Already with 2 or 3 seismometers, residuals are strongly reduced for all plane-wave models. Beyond $N=4$ sensors, the IPW, APW, and SPW curves approximately follow an asymptote proportional to $1/\sqrt{N}$. In other words, 4 sensors are already able to overcome all geometric limitations so that residuals are purely sensor-noise limited. Nonetheless, the absolute sensor noise limit of $1/({\rm SNR}\sqrt{N})$ is not reached by the IPW and APW models independent of the number of sensors. In the case of the IPW, residuals for large numbers of sensors lie about a factor $\sqrt{2}$ above the absolute sensor-noise limit.

In summary, depending on the properties of the seismic field, potential subtraction factors using 6-sensor arrays span 3.5 -- 250, with the APW model representing the best fit to the observed data and corresponding to a suppression by 140. Since inhomogeneities of the seismic field have not been accounted for, it is to be expected that NN residuals achieved in the future, and optimal array designs differ from these predictions. However, the APW model captures the dominant features of the observed correlations, and therefore the methods and results presented here serve at least as first, albeit simplified design of a cancellation system for LIGO (assuming that the seismic fields at the other LIGO stations are not significantly more complex).

\section{Conclusion}
\label{sec:Conclusion}

In this work, we have presented new methods to calculate optimal seismometer-array configurations for Newtonian-noise (NN) cancellation. The results are used to define a first, simplified design of cancellation system applicable to the LIGO detectors.
We showed that cancellation of NN from any type of plane-wave Rayleigh field, i.~e.~with any degree of anisotropy, is very efficient approaching the sensor-noise limit with only a few seismometers. Cancellation performance was investigated in detail for 6-sensor arrays. Only a Gaussian correlation model, which represents a worst-case scenario with respect to NN cancellation, yields insufficient noise cancellation. However, this model is clearly inconsistent with the LIGO Hanford observation, and therefore the upper limit of noise residuals set by this model only serves as worst-case scenario for sites that have not been characterized yet. 

Another requisite for this analysis is that the surface within about a wavelength distance to the test masses is flat. This means specifically that the array designs cannot be readily applied to the Virgo detector where laboratory space below test masses may significantly modify the generation of NN, and of optimal array configurations. In this case, the design of a NN cancellation system must be partially based on numerical simulations to accurately model the effect of seismic scattering and displacement of uneven surfaces.

In the future, we will continue to work on how to find optimal arrays in inhomogeneous fields. Do to the fact that we do not have a method to model inhomogeneous fields, we will continue developing infrastructure to interpolate array data directly instead of fitting to potential models. In addition, we can also consider optimization for underground arrays, which will be limited by NN at the lowest frequencies, and will require inclusion of the Z direction with seismometer placement.

\section{Acknowledgments}
MC was supported by the National Science Foundation Graduate Research Fellowship Program, under NSF grant number DGE 1144152. NM acknowledges Council of Scientific and Industrial Research (CSIR), India for providing financial support as Senior Research Fellow. SM acknowledges the support of the Science and Engineering Research Board (SERB), India through the fast track grant SR/FTP/PS-030/2012.
LIGO was constructed by the California Institute of Technology and Massachusetts Institute of Technology with funding from the National Science Foundation and operates under cooperative agreement PHY-0757058.
This paper has been assigned LIGO document number LIGO-P1600146.

\section*{References}
\bibliographystyle{iopart-num}
\bibliography{references}

\end{document}